\documentclass{spie}
\usepackage{amsmath,amsfonts,amssymb}
\usepackage{graphicx}
\usepackage[colorlinks=true, allcolors=blue]{hyperref}

\begin{document}

\title{Theory of bilinear magneto-electric resistance from topological-insulator surface states}
\author[a,b]{Steven S.-L. Zhang}
\author[b]{Giovanni Vignale}
\affil[a]{Materials Science Division, Argonne National Laboratory, Lemont, Illinois 60439, USA}
\affil[b]{Department of Physics and Astronomy, University of Missouri, Columbia, Missouri 65211, USA}
\authorinfo{Further author information: (Send correspondence to Steven S.-L. Zhang)\\Steven S.-L. Zhang: E-mail: shulei.zhang@anl.gov, Alternative E-mail:  shulei.zhang84@gmail.com}
\maketitle

\begin{abstract}
We theoretically investigate a new kind of nonlinear magnetoresistance on the surface of three-dimensional topological insulators (TIs). At variance with the unidirectional magnetoresistance (UMR) effect in magnetic bilayers, this nonlinear magnetoresistance does not rely on a conducting ferromagnetic layer and scales linearly with both the applied electric and magnetic fields; for this reason, we name it bilinear magneto-electric resistance (BMER). We show that the sign and the magnitude of the BMER depends sensitively on the orientation of the current with respect to the magnetic field as well as the crystallographic axes -- a property that can be utilized to map out the spin texture of the topological surface states via simple transport measurement, alternative to the angle-resolved photoemission spectroscopy (ARPES).
\end{abstract}





\section{INTRODUCTION}

Beyond doubt, the giant magnetoresistance (GMR) effect~\cite%
{Fert88PRL,Grunberg89PRB,Parkin90PRL} is one of the most important
discoveries in the fruitful field of spintronics, which has found various
commercial applications such as magnetic hard disk drives and magnetic
memory devices. The building block of these devices is a trilayer structure
(also known as a spin valve) which consists of two ferromagnetic metal (FM)
layers separated by a nonmagnetic spacer. There is a large variation in the
resistance when the magnetizations of the FM layers switch between parallel and
antiparallel alignments. When a current is passing through
one of the FM layers, the spins of the conduction electrons, due to the
strong exchange coupling with the local magnetic moments, will be polarized
along the magnetization direction of the FM layer (for this reason, this FM
is also called a spin polarizer); the scattering rates of these conduction electrons, when they subsequently propagate
through the other FM layer, will depend on their spin direction relative to the
magnetization orientation of the FM layer, and so will the total resistance of
the spin valve. Therefore, it is the exchange
interaction and the spin-dependent scattering that play the key roles in the
GMR effect.

In the past decade, both theoretical and experimental endeavors have been
dedicated to realizing similar functionalities of a spin valve in magnetic bilayer structures consisting of a ferromagnetic layer and a nonmagnetic
layer with strong spin-orbit coupling. The main idea is to use spin-orbit
coupling together with structural inversion asymmetry to generate net spin
density (or spin accumulation) at the interface through the spin Hall~\cite%
{DYAKONOV71PhysLettA_SHE,Hirsch99PRL,sZhang00PRL,Vignaleg10JSNM,
Sinova15RMP_SHE} or Rashba-Edelstein effect~\cite%
{Rashba84JETPletts,Edelstein90SSC,Tokatly15PRB}. Recently, a small change in
the longitudinal resistance has been observed in several magnetic bilayer
structures~\cite%
{Avci15NatPhys_unidirectionAMR,Jungwirth15PRB_nonlinear-MR-semicond,Avci15APL_AMR,ajFerguson16APL,kjLee17AppPhyExpr_UMR-magnon,Yasuda16PRL_UMR-TI,yxYin&Koopmans17APL_thick-usmr,Lv&jpWang18NatComm_USMR-TI,2018arXiv_magnon-USMR,Pietro2018arXivPRL_USMR}
by reversing the magnetization direction in the presence of an in-plane
current perpendicular to the magnetization, that is to say, $R_{l}\left(
\mathbf{M,j}\right) \neq R_{l}\left( -\mathbf{M,j}\right) $ where $R_{l}$ is
the total longitudinal resistance of the bilayer, $\mathbf{M}$ and $\mathbf{j%
}$ are the magnetization and current-density vectors. Note that by symmetry,
reversing the magnetization in a magnetic layer is equivalent to reversing
the current direction; thus, the magnetoresistance change must be associated
with certain nonlinear current response [for this reason, the
magnetoresistance effect has been coined in the literature the
unidirectional magnetoresistance (UMR)~\cite{Avci15NatPhys_unidirectionAMR}%
], which makes it distinctly different from other linear magnetotransport phenomena
previously studied in magnetic bilayer systems such as the (hybrid) spin Hall
magnetoresistance~\cite{Chien12PRL_Proximity-Pt,Saitoh13PRL_SH-MR,Bauer13PRB_SH-MR,Chien14PRL_AMR}, interfacial
spin-orbit magnetoresistance~\cite{slzhang15PRB_AMR,jXiao14PRB_AMR,slZhang14JAP_AMR}, Hanle magnetoresistance~\cite%
{Dyakonov07PRL_Hanle-MR,Felix16PRL_Hanle-MR,Wu&xfHan16PRB_Hanle-MR},
nonlocal (spin Hall) anomalous Hall~\cite{Bauer13PRB_SH-MR,slZhang16PRL} and
etc. Several different interpretations have been proposed to account for the
UMR effect including the interfacial and bulk spin-dependent scattering
mechanism~\cite%
{Avci15NatPhys_unidirectionAMR,ssZhang&gVignale16PRB_USMR,slZhang17usmr-spie}
and the interfacial spin-flip electron-magnon scattering mechanism~\cite%
{ajFerguson16APL,kjLee17AppPhyExpr_UMR-magnon,Yasuda16PRL_UMR-TI,2018arXiv_magnon-USMR}%
. We note that all these mechanisms rely on one common key ingredient -- the
current-induced interfacial spin accumulation which in turn alters either
the spin asymmetry of the density of the conduction electrons or that of the
scattering rate\footnote{Previous studies have shown that when an in-plane current is applied in a bilayer consisting of a heavy metal and a ferromagnet, spin accumulation will be induced at the interface due to the spin Hall effect in the heavy metal layer. In the presence of the spin-flip electron-magnon scattering, the spin accumulation may create or annihilate interfacial magnons in the ferromagnetic layer, depending on the way it is aligned with the magnetization (either parallel or antiparallel). If the ferromagnetic layer is conducting, such variation in magnon density will in turn alter the scattering rate of the electrons in the ferromagnetic layer and hence change the total resistance of the bilayer accordingly.~\cite{Takahashi10el-magnon,Kajiwara10Nature,sZhang12PRL,sZhang12PRB,vanWees15NatPhys,jShi16natComm_magnon-drag,xfHan16PRB}}.

In this work, we theoretically investigate a new kind of nonlinear
magnetoresistance originating from the topological insulator surface states, which does not
require a magnetic layer. We name this new magnetoresistance effect as
bilinear magneto-electric resistance (BMER) due to its linear scaling with
both the external electric and magnetic fields. At variance with the UMR
effect, the BMER emanates from the conversion of a \textit{nonlinear spin
current} to a charge current rather than the current-induced spin density.
The physical picture of the BMER is schematically depicted in Fig.~\ref%
{fig:BMER-schematics} for the surface states of a TI with hexagonal warping~%
\cite{fLiang09PRL_hexagonal-warping}: Due to the spin-momentum locking of
the topological surface states, electrons in the $\mathbf{k}$ and $-\mathbf{k%
}$ states carry opposite spins and have opposite group velocities [i.e., $%
\mathbf{v}(-\mathbf{k})=-\mathbf{v}(\mathbf{k})$] . When an external
electric field $\mathbf{E}$ is applied along certain $\mathbf{k}$ direction,
its first-order correction to electron distribution is known to be an
accumulation of electrons in the $\mathbf{k}$ states and equal number of
electrons depleted in the $-\mathbf{k}$ states, giving rise to a charge
current $\mathbf{j}_{e}=\sigma \mathbf{E}$ and a net spin density $\delta
\mathbf{s\sim \hat{z}\times E}$ with $\mathbf{\hat{z}}$ denoting the normal
vector perpendicular to the surface, while the second-order correction to
the electron distribution, which has not been well studied, results in equal number of electrons populated
in the surface states with opposite momenta as well as spins and thus
induces a \textit{nonlinear pure spin current} $\mathbf{j}_{s}\sim \mathbf{E}%
^{2}$\footnote{Similar nonlinear spin current was also proposed~\cite{Nagaosa17PRB_Nonlinear-spin-current,Tokura17Nat.Phys_recification-semicond} recently in other systems with broken inversion symmetry.}. When a magnetic field is applied, both the group velocity and the
second-order distribution are shifted in $\mathbf{k}$-space and then the two
fluxes of electrons with opposite spin orientations no longer compensate
each other, causing the spin current to be partially converted into a charge
current $\delta \mathbf{j}_{e}\sim \mathbf{E}^{2}$. From an application
perspective, the BMER can be used to map the spin texture of surface states
with spin-momentum locking by simple transport measurements, which has been
demonstrated experimentally on the conducting surface of the topological
insulator Bi$_{2}$Se$_{3}$~\cite{HeZhang18NP_BMER} and for the two dimensional electron gas on the (111) surface
of SrTiO$_{3}$~\cite{hPan18PRL_STO-BMER}.

\begin{figure}[h]
\centering
\includegraphics[trim={2cm 1cm 2cm 2cm},clip=true,
width=0.8\linewidth]{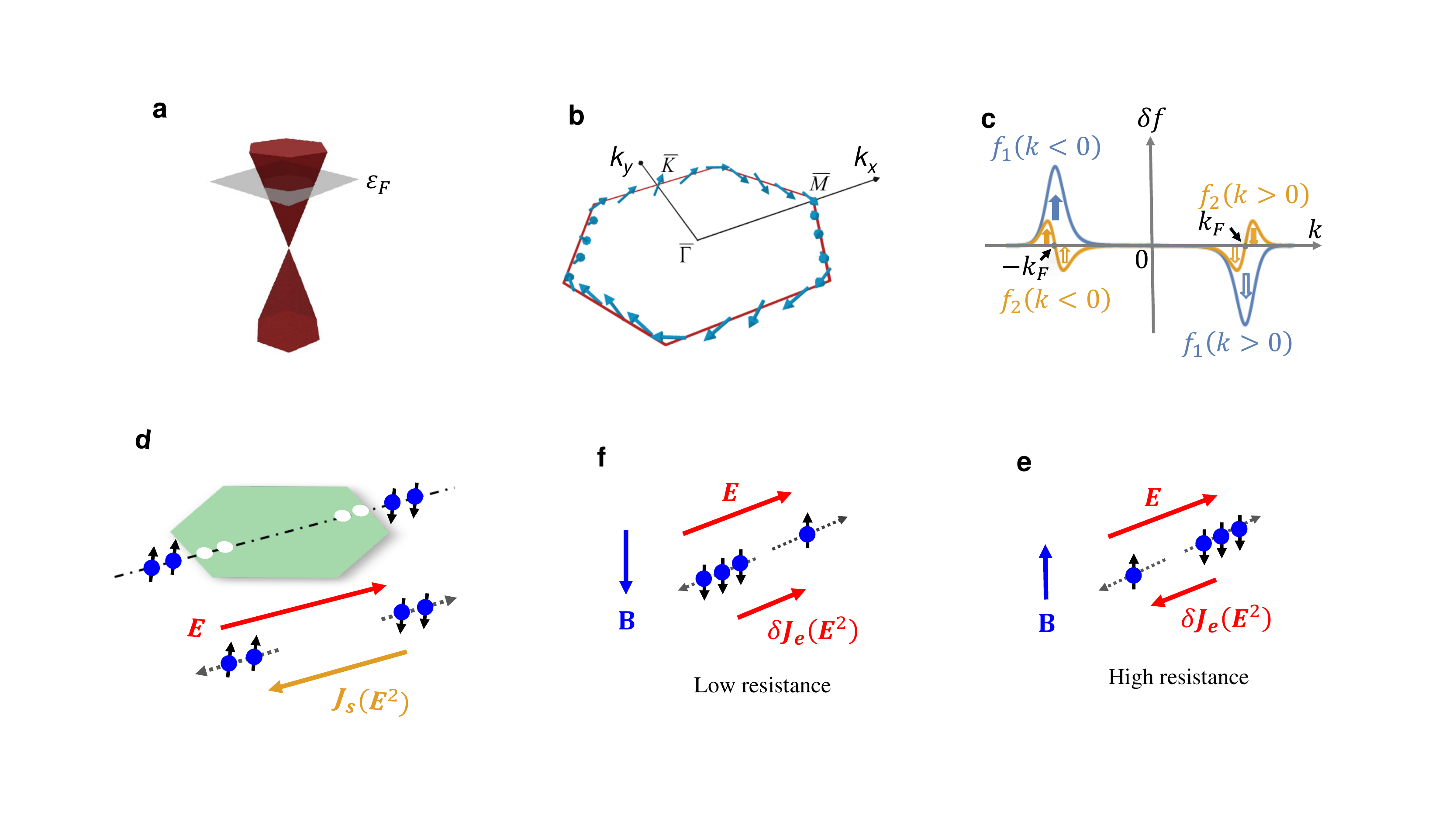}
\caption{Schematics of the physical mechanism of the BMER arising from the TI
surface states with hexagonal warping. Panel a: Dirac cone in a 3D TI with
hexagonal warping. Panel b: Spin texture of the Fermi contour in the
presence of hexagonal warping. Panel c: Variation of the electron
distribution in the applied electric field $\mathbf{E}$: $f_{1}$ (blue curve) and $f_{2}$ (yellow curve) are
respectively the corrections to the equilibrium distribution of the first-
and second-order in $\mathbf{E}$. Solid arrows represent excess of electrons
with spins along the arrow direction, and hollow arrows represent depletion
of the same. Panel d: When an external electric field $\mathbf{E}$ is
applied along a certain direction in the momentum space (dash-dotted line),
a nonlinear pure spin current $\mathbf{j}_{s}(\mathbf{E}^{2})$ is generated
at the second order in $\mathbf{E}$, due to spin--momentum locking. When an
external magnetic field $\mathbf{B}$ is applied, the nonlinear spin current
is partially converted into a charge current $\delta\mathbf{j}_{e}(\mathbf{E}^{2})$%
: a high(low)-resistance state can be reached by applying a magnetic field
antiparallel (parallel) to the spin direction of the electronic states with $%
\mathbf{k}\parallel \mathbf{E}$, as shown in panels e and f respectively. }
\label{fig:BMER-schematics}
\end{figure}

The remainder of the paper is organized as follows. The general formulation
of the nonlinear current response of the TI is developed in Sec.~\ref%
{Sec:general-formula}: We start with the model Hamiltonian for the TI surface states in Sec.~\ref{Sec:Hamiltonian} followed by a discussion of
general symmetry considerations of the system and the nonlinear current-response
function in Sec.~\ref{Sec:symmetry}; then we solve the semi-classical
Boltzmann equation up to the second order in the external electric field in
Sec.~\ref{Sec:Boltzmann-eq} and present the detailed derivation of the nonlinear
current response function in Sec.~\ref{Sec:derv-response}. In Sec.~\ref%
{Sec:BMER}, we use the general formula for the nonlinear response function
to derive an analytical expression for the BMER from the TI surface
states in the presence of hexagonal warping, and then provide more detailed
discussions on various aspects of the BMER effect, including its dependences
on the hexagonal warping (Sec.~\ref{Sec:BMER-lambda}), the momentum relaxation
time (Sec.~\ref{Sec:BEMR-tau}) as well as the Berry curvature effect (Sec.~\ref%
{Sec:Berry}). Finally, we summarize our main results in Sec.~\ref{summary}.

\bigskip

\label{sec:intro} 

\section{GENERAL FORMULATION}

\label{Sec:general-formula}

\subsection{Model Hamiltonian}

\label{Sec:Hamiltonian} Let us start with the following model Hamiltonian
for a Dirac electron in the topological surface state:

\begin{equation}
\mathcal{H}_{TI}=\boldsymbol{\sigma }\cdot \left[ \mathbf{h}\left( \mathbf{k}%
\right) +g\mu _{B}\mathbf{\mathbf{B}}\right]  \label{Eq:H_TI}
\end{equation}%
with $\boldsymbol{\sigma }$ the Pauli spin matrices, $\mathbf{\mathbf{B}}$
the uniform external magnetic field, $g$ and $\mu _{B}$ representing the $g$%
-factor and the Bohr magneton respectively, and
\begin{equation}
\mathbf{h}\left( \mathbf{k}\right) =\alpha \hbar \mathbf{k\times \hat{z}+}%
\lambda \mathbf{k\times \hat{y}}\left( k_{x}^{2}-3k_{y}^{2}\right) \,,
\label{Eq:h(k)}
\end{equation}%
where $\alpha $ is the Dirac velocity, and the term cubic-in-$k$ describes the hexagonal warping effect~\cite%
{fLiang09PRL_hexagonal-warping}. Note that the hexagonal warping
term leads to a threefold rotational symmetry $C_{3v}$, which becomes more
transparent when we rewrite $\mathbf{h}\left( \mathbf{k}\right) $ in its
angular form as
\begin{equation}
\mathbf{h}\left( \mathbf{k}\right) =\alpha \hbar k\left( \sin \phi _{k}%
\mathbf{\hat{x}-}\cos \phi _{k}\mathbf{\hat{y}}\right) \mathbf{+}\left(
\frac{1}{2}\lambda k^{3}\cos 3\phi _{k}\right) \,\mathbf{\hat{z}} \,,
\end{equation}%
where $\phi _{k}$ is the azimuthal angle of the wavevector $\mathbf{k}$ with
respect to the $x$-axis.

The energy dispersion can be obtained by
\begin{equation}
\varepsilon ^{s}\left( \mathbf{k}\right) =s\left\vert \mathbf{h}\left(
\mathbf{k}\right) +g\mu _{B}\mathbf{\mathbf{B}}\right\vert \,,
\label{Eq:E^s}
\end{equation}%
where $s=+1$ and $-1$ correspond to the upper and lower surface bands
respectively, and the group velocity is given by
\begin{equation}
\mathbf{v}^{s}\left( \mathbf{k}\right) =\frac{\partial \varepsilon
^{s}\left( \mathbf{k}\right) }{\hbar \partial \mathbf{k}} \,.
\end{equation}%
Note that the group velocity is odd in $\mathbf{k}$ in the absence of the
external magnetic field. In what follows, we shall assume the Fermi level
lies in the upper band ($s=+1$) and thus suppress the superscript for the
band index hereafter. Also note that the Berry curvature effects on
the density of states and the orbital magnetic moment are not included since they are negligibly small when the Fermi surface is far away from the Dirac point, as we will show in Sec.~\ref{Sec:Berry}.

\subsection{Symmetry considerations}

\label{Sec:symmetry} In the absence of the external magnetic field, the
effective Hamiltonian for the topological surface states, i.e., $\mathcal{H}%
_{TI}^{\left( 0\right) }=\boldsymbol{\sigma }\cdot \mathbf{h}\left( \mathbf{k%
}\right) $, is invariant under the following two operations: 1) Mirror
reflection about the $y$-$z$ plane, i.e., $M:$ $x\rightarrow -x$, and 2)
threefold rotation $C_{3}$ about the $z$-axis. Here, we are interested in the
nonlinear current-density response to the second-order electric field and
the first-order magnetic field, i.e.,
\begin{equation}
j_{e,a}^{(2)}=\sum_{bcd}K_{abcd}E_{b}E_{c}B_{d}  \label{Eq: j_a^(2)_v0}  \,,
\end{equation}%
where the response function $K_{abcd}$ is a fourth-rank tensor with indices $a,b,c=x$ or $y$ and $d=x,y$ or $z$. For the
transport in the surface states, the external electric field is applied in
the $x$-$y$ plane (parallel to the surface of the TI) while the magnetic field
can be three-dimensional; therefore, the tensor $K_{abcd}$ has 24 elements in
total. As the response function is the property of the unperturbed system,
we should expect it to reflect the same symmetries as those of $\mathcal{H}%
_{TI}^{\left( 0\right) }$.

First, we note that the mirror symmetry of the unperturbed 2D system
requires that the tensor element $K_{abcd}$\ must be zero if, under mirror
reflection operation $M:$ $x\rightarrow -x$, the corresponding current
component $j_{e,a}$\ changes sign whereas the product of the external fields $%
E_{b}E_{c}H_{d}$\ is invariant, and vice versa. Without doing any further
calculation, we know the following 12 tensor elements are zero, i.e.,
\begin{equation*}
K_{xxxx}=K_{xxyy}=K_{xyxy}=K_{xyyx}=K_{xxyz}=K_{xyxz}=0
\end{equation*}%
and
\begin{equation*}
K_{yyyy}=K_{yyxx}=K_{yxyx}=K_{yxxy}=K_{yyyz}=K_{yxxz}=0 \,.
\end{equation*}
The other 12 tensor elements could remain finite. Furthermore, arising from
the threefold rotational symmetry about the $z$-axis, some of the remaining 12
tensor elements have equal values, as we will show explicitly below.

\subsection{Second-order nonequilibrium distribution function}

\label{Sec:Boltzmann-eq} Now let us examine the following single-band
steady-state Boltzmann equation
\begin{equation}
\frac{E_{a}}{\hbar }\frac{\partial f}{\partial k_{a}}=-\frac{f-f_{0}}{\tau }  \,,
\label{Eq:BTE}
\end{equation}%
where we have assumed, for simplicity, a constant relaxation time $\tau$. Expand
the distribution function in powers of the electric field, i.e.,
\begin{equation}
f=f_{0}+f_{1}+f_{2}+...  \label{Eq:f-expansion} \,,
\end{equation}%
where $f_{0}$ is the equilibrium Fermi distribution, $f_{1}$ and $f_{2}$ are
the nonequilibrium distribution functions of the first- and second-order in
the electric field, i.e., $f_{1}\propto $ $E_{a}$ and $f_{2}\propto
E_{a}E_{b}$. Placing the expansion~(\ref{Eq:f-expansion}) in the Boltzmann
equation (\ref{Eq:BTE}) and equating terms on the left and right sides of equal order in
the electric field, we get
\begin{equation}
\frac{E_{a}}{\hbar }\frac{\partial f_{i}}{\partial k_{a}}=-\frac{f_{i+1}}{%
\tau }  \,,
\end{equation}%
where $i=0,1,2,...$ By solving the series of equations iteratively, we find
the first-order nonequilibrium distribution function takes the familiar form
\begin{equation}
f_{1}=-\frac{\tau E_{a}}{\hbar }\frac{\partial f_{0}}{\partial k_{a}}
\end{equation}%
and the second-order nonequilibrium distribution function of interest can
be expressed as
\begin{equation}
f_{2}=\frac{\tau ^{2}}{2\hbar ^{2}}\sum_{ab}\frac{\partial ^{2}f_{0}}{%
\partial k_{a}\partial k_{b}}E_{a}E_{b}  \,,
\end{equation}%
where the prefactor $\frac{1}{2}$ eliminates double counting in the
summation.

\subsection{Nonlinear current response function}

\label{Sec:derv-response} The charge current density can be calculated via $\mathbf{%
j}_e=-e\int \frac{d^{2}\mathbf{k}}{\left( 2\pi \right) ^{2}}\mathbf{v}\left(
\mathbf{k}\right) f\left( \mathbf{k}\right) $. It follows that the nonlinear component of the
current density of interest can be expressed as
\begin{equation}
j_{e,a}^{(2)}=-eg\mu _{B}\left( \frac{e\tau }{\hbar }\right) ^{2}\frac{1}{2}%
\sum_{\mathbf{k},bcd}\left( \frac{\partial ^{3}f_{0}}{\partial k_{b}\partial
k_{c}\partial h_{d}}v_{a}+\frac{\partial ^{2}f_{0}}{\partial k_{b}\partial
k_{c}}\frac{\partial v_{a}}{\partial h_{d}}\right) E_{b}E_{c}B_{d}  \,,
\label{Eq:j_a^(2)}
\end{equation}%
where we have used the relation $\left. \frac{\partial \varepsilon }{%
\partial B_{d}}\right\vert _{\mathbf{B\rightarrow 0}}=g\mu _{B}\frac{%
\partial \varepsilon }{\partial h_{d}}$ with $h_{d}$ ($d=x,y$ or $z$) the
Cartesian components of the vector $\mathbf{h(k)}$ given by Eq.~(\ref%
{Eq:h(k)}). Note that all the derivatives in Eq.~(\ref{Eq:j_a^(2)}) are
calculated at zero external magnetic field. Comparing Eq.~(\ref{Eq:j_a^(2)})
with Eq.~(\ref{Eq: j_a^(2)_v0}), we identify the nonlinear current response
function tensor as%
\begin{equation}
K_{abcd}=-eg\mu _{B}\left( \frac{e\tau }{\hbar }\right) ^{2}\frac{1}{2}\sum_{%
\mathbf{k}}\left( \frac{\partial ^{3}f_{0}}{\partial k_{b}\partial
k_{c}\partial h_{d}}v_{a}+\frac{\partial ^{2}f_{0}}{\partial k_{b}\partial
k_{c}}\frac{\partial v_{a}}{\partial h_{d}}\right) \,.
\end{equation}%
Performing integration by parts on the r.h.s. of the above equation, we get
\begin{equation}
K_{abcd}=-eg\mu _{B}\left( \frac{e\tau }{\hbar }\right) ^{2}\frac{1}{4}\sum_{%
\mathbf{k}}f_{0}^{\prime }\left( \frac{\partial h}{\partial h_{d}}\frac{%
\partial ^{2}v_{a}}{\partial k_{b}\partial k_{c}}-\frac{\partial h}{\partial
k_{b}}\frac{\partial ^{2}v_{a}}{\partial k_{c}\partial h_{d}}+\left\{
b\leftrightarrow c\right\} \right)  \,,
\end{equation}%
where $f_{0}^{\prime }\equiv \frac{\partial f_{0}}{\partial h}$ with $%
h\equiv \left\vert \mathbf{h(k)}\right\vert $, and $\left\{ b\leftrightarrow
c\right\} $ is the shorthand notation of two more terms that are simply the first
two terms in the parentheses with the indices $b$ and $c$ interchanged. Note that when the temperature at which the current is measured is much lower
than the Fermi temperature, it is a good approximation to replace $%
f_{0}^{\prime }\ $with the delta function $-\delta \left( h-\varepsilon
_{F}\right) $ and it follows that $\int d^{2}\mathbf{k}\delta \left(
h-\varepsilon _{F}\right) \frac{F\left( \mathbf{k}\right) }{h}%
=\oint_{FL}kd\phi _{k}\frac{2F\left( \mathbf{k}\right) }{|\nabla _{k}h^{2}|}$.
It is straightforward to express the derivatives in the parentheses as those of $h^{2}$ or $%
h_{a}$ with respect to $k_{b}$ ($a,b=x,y,$or $z$) as follows
\begin{eqnarray}
K_{abcd} &=&\frac{e^{3}\tau ^{2}g\mu _{B}}{8\pi ^{2}\hbar ^{3}}\oint_{FL}%
\frac{kd\phi _{k}}{|\nabla _{k}h^{2}|}\left\{ \frac{\partial _{abc}h^{2}}{h}-%
\frac{2(\partial _{a}h^{2})(\partial _{bc}h^{2})+(\partial
_{b}h^{2})(\partial _{ca}h^{2})+(\partial _{c}h^{2})(\partial _{ba}h^{2})}{%
4h^{3}}\right\} h_{d}  \notag \\
&&-\frac{e^{3}\tau ^{2}g\mu _{B}}{8\pi ^{2}\hbar ^{3}}\oint_{FL}\frac{kd\phi
_{k}}{|\nabla _{k}h^{2}|}\frac{(\partial _{ab}h^{2})(\partial
_{c}h_{d})+(\partial _{ac}h^{2})(\partial _{b}h_{d})}{2h}  \notag \\
&&+\frac{e^{3}\tau ^{2}g\mu _{B}}{8\pi ^{2}\hbar ^{3}}\oint_{FL}\frac{kd\phi
_{k}}{|\nabla _{k}h^{2}|}\frac{2(\partial _{a}h_{d})(\partial
_{b}h^{2})(\partial _{c}h^{2})+(\partial _{b}h_{d})(\partial
_{c}h^{2})(\partial _{a}h^{2})+(\partial _{c}h_{d})(\partial
_{a}h^{2})(\partial _{b}h^{2})}{4h^{3}}  \notag  \,, \\
&&  \label{Eq:K_abcd-final}
\end{eqnarray}%
where we have used the identities $\frac{\partial h}{\partial h_{d}}=\frac{%
h_{d}}{h}$ and $\partial _{i}h\equiv \frac{\partial h}{\partial k_{i}}=\frac{%
\partial _{i}h^{2}}{2h}$, and have converted the summation over $\mathbf{k}$
to an integral over the Fermi loop (FL) where $h=\varepsilon _{F}$ is a
constant.

\section{RESULTS AND DISCUSSIONS}

\subsection{Bilinear magneto-electric resistance (BMER)}

\label{Sec:BMER} Inserting Eq.~(\ref{Eq:h(k)}) into Eq.~(\ref%
{Eq:K_abcd-final}), we can calculate all the elements of the nonlinear
current-response function tensor. The results for the nonzero tensor
elements are
\begin{equation*}
K_{xyyz}=K_{yxyz}=K_{yyxz}=-K_{xxxz}=\frac{3\kappa _{0}}{8\pi }\cdot \frac{%
\lambda \varepsilon _{F}}{\left( \alpha \hbar \right) ^{2}}
\end{equation*}%
\begin{equation*}
K_{yxxx}=K_{xxyx}=K_{xyxx}=-K_{xyyy}=-K_{yxyy}=-K_{yyxy}=\frac{1}{3}%
K_{yyyx}=-\frac{1}{3}K_{xxxy}=\frac{3\kappa _{0}}{4\pi }\cdot \frac{\lambda
^{2}\varepsilon _{F}^{3}}{\left( \alpha \hbar \right) ^{5}}  \,,
\end{equation*}%
where $\kappa _{0}=g\mu _{B}e^{3}\tau ^{2}/\hbar ^{3}$. All other 12 tensor
elements are zero, in agreement with the symmetry analysis that we carried
out earlier. Note that in deriving these results, we have assumed
that the linear term in the Hamiltonian for the TI surface states~(\ref{Eq:H_TI}), giving rise to the spin-momentum locking, is dominant over the
cubic hexagonal warping term.

Having known these tensor elements, we can write the two Cartesian
components of the current density as
\begin{subequations}
\begin{equation}
j_{e,x}^{\left( 2\right) }=-c_{\parallel }E_{x}^{2}B_{y}+\frac{2}{3}%
c_{\parallel }E_{x}E_{y}B_{x}-\frac{1}{3}c_{\parallel
}E_{y}^{2}B_{y}-c_{\perp }\left( E_{x}^{2}-E_{y}^{2}\right) B_{z}
\end{equation}%
\begin{equation}
j_{e,y}^{\left( 2\right) }=c_{\parallel }E_{y}^{2}B_{x}-\frac{2}{3}%
c_{\parallel }E_{x}E_{y}B_{y}+\frac{1}{3}c_{\parallel
}E_{x}^{2}B_{x}+2c_{\perp }E_{x}E_{y}B_{z}  \,,
\end{equation}%
\end{subequations}
where $c_{\parallel }=\frac{9\lambda ^{2}e^{3}\tau ^{2}g\mu _{B}\varepsilon
_{F}^{3}}{4\pi \alpha ^{5}\hbar ^{8}}$ and $c_{\perp }=\frac{3\lambda
e^{3}\tau ^{2}g\mu _{B}\varepsilon _{F}}{8\pi \alpha ^{2}\hbar ^{5}}$
with the subscripts ``$\parallel$" and ``$\perp $
" refer to the terms associated with the in-plane and out-of-plane
components of the external magnetic field respectively. The longitudinal
resistivity can be calculated via $\rho _{l}=\mathbf{E\cdot j}_e/\left\vert
\mathbf{j}_e\right\vert ^{2}$ with $\mathbf{j}_e (=\mathbf{j}_e^{\left( 1\right) }+\mathbf{j}_e%
^{\left( 2\right) }+...$) the total current density. Up to the first order in the
external electric field, we obtain

\begin{equation}
\rho _{l}=\rho _{0}-E\left[ \rho _{\parallel }^{\left( 2\right) }\left(
B_{x}\sin \phi _{E}-B_{y}\cos \phi _{E}\right) -\rho _{\perp }^{\left(
2\right) }B_{z}\cos 3\phi _{E}\right] +O\left( \mathbf{E}^{2}\right) \,,
\label{Eq:rho_l}
\end{equation}%
where $E$ is the magnitude of the external electric field $\mathbf{E}$, $%
\phi _{E}$ is the angle between $\mathbf{E}$ and the $x$-axis,$\,\rho _{0}=%
\frac{4\pi \hbar ^{2}}{\tau e^{2}\varepsilon _{F}}$ is the linear surface
resistivity independent of $\mathbf{E}$, and the coefficients
\begin{equation}
\rho _{\parallel }^{\left( 2\right) }=\left( \frac{36\pi g\mu _{B}}{e\hbar
^{4}}\right) \frac{\lambda ^{2}\varepsilon _{F}}{\alpha ^{5}}\text{ and }%
\rho _{\bot }^{\left( 2\right) }=\left( \frac{6\pi g\mu _{B}}{e\hbar }%
\right) \frac{\lambda }{\alpha ^{2}\varepsilon _{F}}  \label{Eq:rho^(2)}
\end{equation}%
characterize the magnitudes of the nonlinear surface resistivities for the
in-plane and out-of-plane components of the external magnetic fields respectively.

Now we are in a position to provide some remarks on the nonlinear component of the surface resistivity,
i.e, the second term on the r.h.s. of Eq.~(\ref{Eq:rho_l}): 1) The nonlinear
resistivity is linearly proportional to the electric and magnetic fields. 2)
The nonlinear resistivity is inversely proportional to $e$, similar to the
regular Hall coefficient; thus one would expect it to change sign as the
type of the charge carrier changes from electrons to holes, and vice versa. 3)
The nonlinear resistivities associated with the in-plane and out-of-plane components of the magnetic
fields exhibit different dependences on the hexagonal warping, i.e., $\rho
_{\parallel }^{\left( 2\right) }\propto \lambda ^{2}$ and $\rho _{\bot
}^{\left( 2\right) }\propto \lambda $, and both of them vanish when $\lambda
\rightarrow 0$, because the in-plane and out-of-plane magnetic fields play
different roles in altering the energy dispersion of the surface states. We
will elaborate on this point in Sec.~\ref{Sec:BMER-lambda}. 4) As shown by
Eq.~(\ref{Eq:rho^(2)}), the nonlinear resistivity turns out to be
independent of the relaxation time constant $\tau $ and only depends on the
main material parameters of the TI surface states (namely $\alpha $, $\lambda $ and $%
\varepsilon _{F}$). This is somewhat surprising as it indicates that the
nonlinear resistivity is independent of scatterings. We note that this
finding should be taken with caution due to the constant relaxation time
approximation made in our model calculation; we will discuss in more details
the validity of the approximation in Sec.~\ref{Sec:BEMR-tau}. 5) In addition
to the longitudinal component of the nonlinear resistivity, one can also
derive its transverse counterpart via $\rho _{t}=\mathbf{\hat{z}\cdot }%
\left( \mathbf{E}\times \mathbf{j}_e\right) /\left\vert \mathbf{j}_e\right\vert ^{2}$; in
order to concentrate on the BMER effect, we will study this nonlinear Hall
effect elsewhere~\cite{HeZhang_nlHall_unpub}.

\begin{figure}[h]
\centering
\includegraphics[trim={0.5cm 8cm 0.8cm 4cm},clip=true,
width=0.9\linewidth]{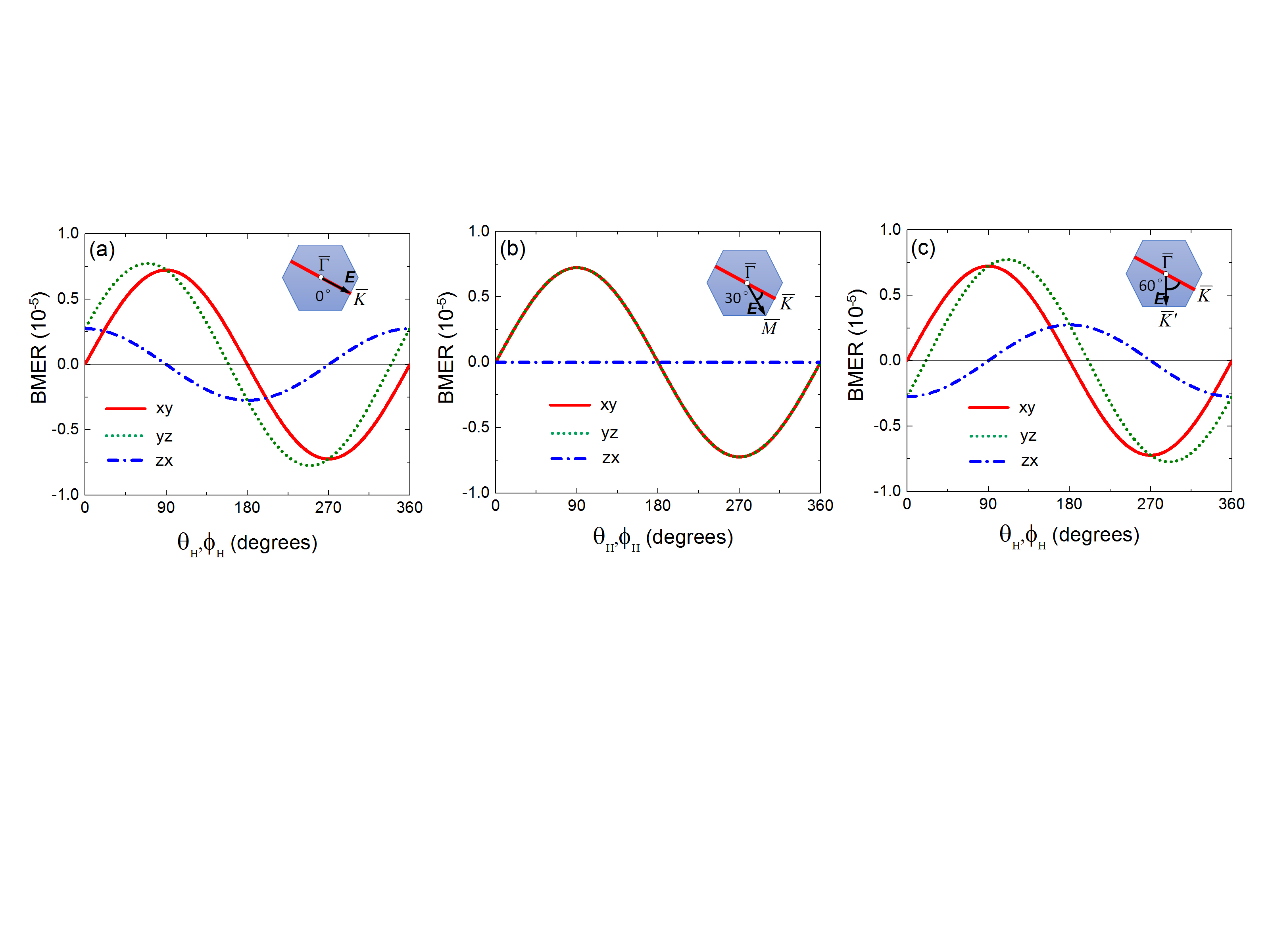}
\caption{Angular dependences of the BMER for magnetic field swept in the x-y,
y-z and z-x planes and with electric field applied along three typical
crystallographic axes: (a) $\mathbf{E}\parallel \overline{\Gamma K}$ ($%
\protect\phi_E=0^{\circ}$), (a) $\mathbf{E}\parallel \overline{\Gamma M}$ ($%
\protect\phi_E=30^{\circ}$), and (c) $\mathbf{E}\parallel \overline{\Gamma
K^{\prime }}$ ($\protect\phi_E=60^{\circ}$). Note that in our
coordinate system, $\overline{\Gamma K}$ is along the $x$-axis. Parameters
used: $\protect\alpha=5 \times 10^5$ m$\cdot$s$^{-1}$, $\protect\lambda =165$
eV$\cdot $\AA $^{3}$, $\protect\varepsilon_{F}=0.256$ eV, $g=2$, $B=9$ T and
$E=100$ V$\cdot$cm$^{-1}$.}
\label{fig:BMER-angles}
\end{figure}

To evaluate the magnitude of the nonlinear resistivity relative to the
linear resistivity $\rho _{0}$, we may define the BMER as follows
\begin{equation}
BMER\equiv \frac{\rho _{l}\left( \mathbf{E},\mathbf{B}\right) -\rho
_{l}\left( -\mathbf{E},\mathbf{B}\right) }{\rho _{l}\left( \mathbf{E},%
\mathbf{B}\right) +\rho _{l}\left( -\mathbf{E},\mathbf{B}\right) }  \,.
\label{Eq:BMER-def}
\end{equation}%
Placing Eq.~(\ref{Eq:rho_l}) in (\ref{Eq:BMER-def}) and expanding terms up
to the first order in $\mathbf{E}$, we get
\begin{equation}
BMER=\frac{EB}{\rho _{0}}\left[ \rho _{\parallel }^{\left( 2\right) }\sin
\theta _{H}\sin \left( \phi _{H}-\phi _{E}\right) +\rho _{\perp }^{\left(
2\right) }\cos \theta _{H}\cos 3\phi _{E}\right] \,,
\end{equation}%
where $\theta _{H}$ and $\phi _{H}$ are, respectively, the polar and
azimuthal angles of the magnetic field, and $B$ is the magnitude of the
magnetic field. In Fig.~\ref{fig:BMER-angles}, we show the angular
dependence of the BMER for magnetic field scans in $x$-$y$, $y$-$z$ and $z$-$x$ planes with the electric field applied along three typical crystallographic axes.

\subsection{Dependence of BMER on the hexagonal warping}

\label{Sec:BMER-lambda} As we noted earlier that the BMER's for the in-plane
and out-of-plane magnetic fields exhibit different dependences on the
hexagonal warping, i.e., $\rho _{\parallel }^{\left( 2\right) }\propto
\lambda ^{2}$ and $\rho _{\bot }^{\left( 2\right) }\propto \lambda $. The
two distinct dependences of BMER on the hexagonal warping emanate from the
different roles played by the in-plane and out-of-plane magnetic fields to
the surface band structure of the TI. This can be seen by rewriting Eq.~(\ref%
{Eq:E^s}) as follows
\begin{equation}
\varepsilon ^{s}\left( \mathbf{k}\right) =s\sqrt{\left[ \alpha \hbar \left(
\mathbf{k-}\delta \mathbf{k}\right) \mathbf{\times \hat{z}}\right] ^{2}+%
\left[ \lambda k^{3}\cos \left( 3\phi _{k}\right) +\Delta_{g}\right] ^{2}%
}  \,,
\end{equation}%
where
\begin{equation}
\delta \mathbf{k=}\frac{g\mu _{B}}{\alpha \hbar }\mathbf{B}_{\parallel }%
\mathbf{\times \hat{z}}\text{ and }\Delta_{g}=g\mu _{B}B_{z}  \,.
\end{equation}%
In the absence of the hexagonal warping effect (i.e, $\lambda =0$), we can
see that an in-plane magnetic field shifts the Dirac cone rigidly in the $%
k_{x}$-$k_{y}$ plane, whereas an out-of-plane magnetic field opens up a band
gap of $2\Delta_{g}$ at the Dirac point. Consequently, without the cubic
hexagonal warping term, applying an in-plane magnetic field would not alter
the current provided the system has translational symmetry in the $x$-$y$
plane, and likewise an out-of-plane magnetic field wouldn't do so as long as
the Fermi level lies far away from the gap opened by $B_{z}$ (note that the
gap is about $2$ meV for $B_{z}=10$ T and $g=2$).

In the presence of the hexagonal warping effect, however, a magnetic field
not only shifts the Dirac cone and/or open up a gap, but also deforms the
snow-flake-like Fermi contour accordingly, as shown in Fig.~\ref%
{fig:Dirac-cone-B}. It follows that both the group velocity and the
second-order distribution function are shifted in $\mathbf{k}$-space in a
way that the two fluxes of electrons with opposite spin orientations no
longer compensate each other, giving rise to the BMER effect.

\begin{figure}[h]
\centering
\includegraphics[trim={0.0cm 0cm 0.0cm 0cm},clip=true,
width=0.9\linewidth]{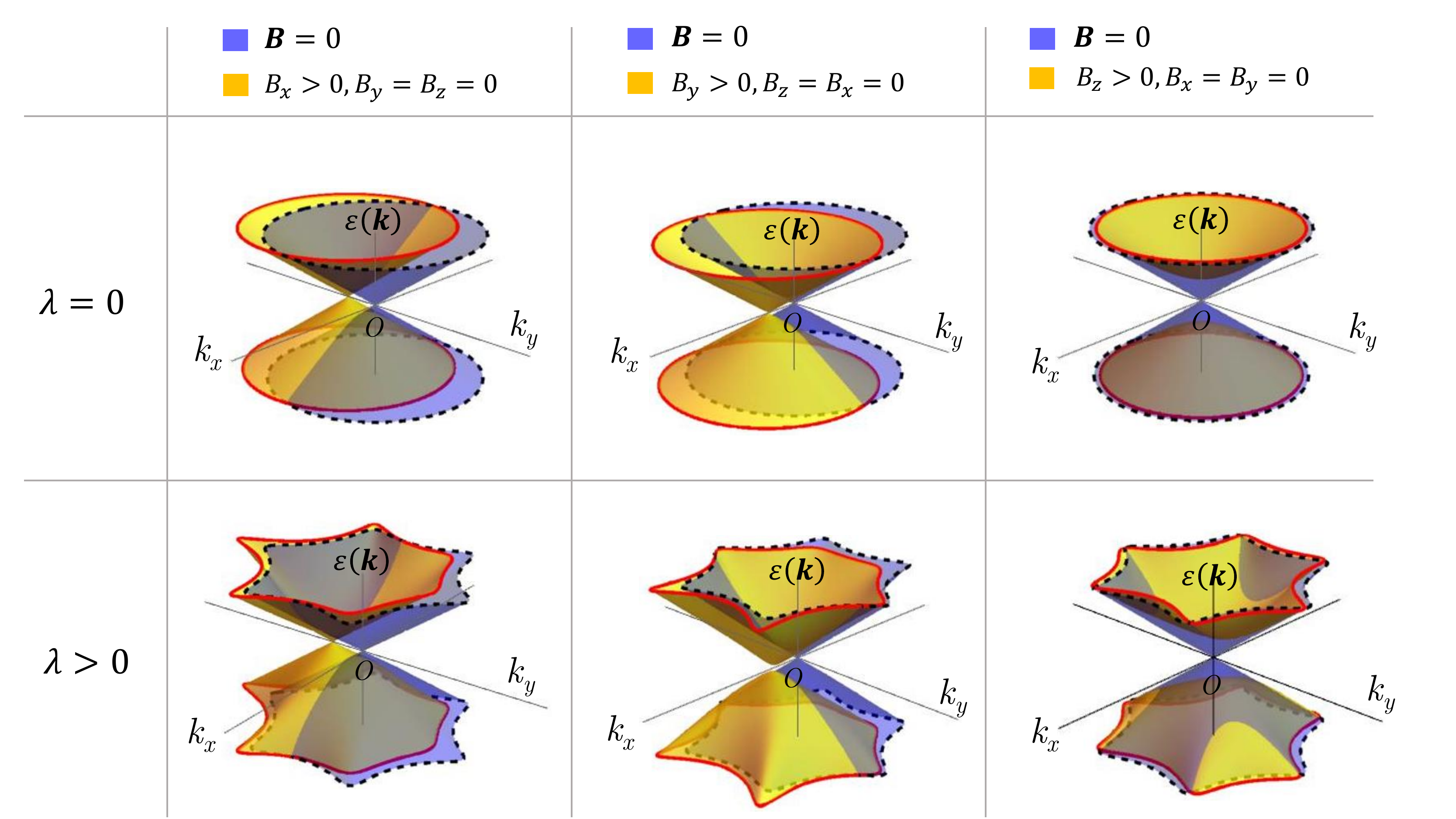}
\caption{Magnetic-field induced deformation of the surface energy dispersion
$\protect\varepsilon(\mathbf{k})$. The top panels show the Dirac cones in
the absence of the hexagonal warping ($\protect\lambda=0$), and the bottom
panels show those in the presence of the hexagonal warping ($\protect\lambda %
\neq 0$). In each panel, the Dirac cone in the absence of the
external magnetic field is depicted in blue color, whereas that in the presence of
the external magnetic field is depicted in orange color.}
\label{fig:Dirac-cone-B}
\end{figure}

\subsection{Dependence of BMER on the momentum relaxation time}

\label{Sec:BEMR-tau} As we have pointed out in the end of Sec.~\ref{Sec:BMER}%
, the nonlinear longitudinal resistivity turns out to be independent of the
relaxation time $\tau$. Here we want to emphasize that this result relies on
two approximations that we made in our calculation: 1) constant relaxation
time approximation and 2) single-band contribution to the conductivity. The
constant relaxation time approximation is valid when the scattering
potential is isotropic and short-ranged. However, when there are multiple
bands contributing to the conduction, the nonlinear resistivity would still
depend on the relaxation time of each band even if the relaxation times
are momentum independent. To see this, let us consider a simple case of two
bands with constant relaxation times $\tau _{a}$ and $\tau _{b}$. The total
current density is the sum of the contributions from the two bands
(neglecting interband transition); for a simple 1-D problem in which an
electric field is applied in the x-direction, the total current
density is given by%
\begin{equation}
j_{x}=j_{a,x}+j_{b,x}=(c_{a,1}\tau _{a}+c_{b,1}\tau _{b})E_{x}+(c_{a,2}\tau
_{a}^{2}+c_{b,2}\tau _{b}^{2})E_{x}^{2}+O(E_{x}^{3}) \,,
\end{equation}%
where $c_{a(b),1}$ and $c_{a(b),2}$ are coefficients independent of $\tau
_{a}$, $\tau _{b}$ and $E_{x}$. The resistivity can be obtained
by $\rho _{xx}=E_{x}/j_{x}$ ; up to the 2nd order in $E_{x}$, we get,

\begin{equation}
\rho _{xx}\cong \frac{1}{c_{a,1}\tau _{a}+c_{b,1}\tau _{b}}\left[ 1-\left(
\frac{c_{a,2}\tau _{a}^{2}+c_{b,2}\tau _{b}^{2}}{c_{a,1}\tau
_{a}+c_{b,1}\tau _{b}}\right) E_{x}\right] +O(E_{x}^{2})  \,.
\end{equation}%
We see that the $\tau$-dependence of $\rho _{xx}$ disappears in
the single-band case when either $c_{a,i}$ or $c_{b,i}$ ($i=1,2$) are set to
zero; however, the nonlinear resistivity (i.e., the term linear in $E_{x}$)
in general depends on the relaxation times in the two-band case when both $%
c_{a,i}$ and $c_{b,i}$ remain finite. The same conclusion holds for
multiple-band ($i>3$) cases.

\subsection{Influence of the Berry phase effect on BMER}

\label{Sec:Berry}Hitherto we have not considered the influence of the Berry
curvature of the surface bands on the nonlinear transport, which we shall
justify below. Qualitatively speaking, the BMER effect relies on the
hexagonal warping term which becomes important when the Fermi level lies far
away from the Dirac point, as we have discussed in Sec.~\ref{Sec:BMER-lambda}%
, whereas the Berry curvature effect is profound when the Fermi level is
close to Dirac point. For the Bi$_2$Se$_3$ investigated in the recent
experiment~\cite{HeZhang18NP_BMER}, the Fermi level lies far away from the
Dirac point, and hence the Berry curvature effect on the BMER is negligible. Below, we will perform an order-of-magnitude estimation to
confirm this.

In the presence of an external magnetic field, the Berry curvature alters
the transport property in two ways: 1) a correction to the density of states
~\cite{XiaoNiu05PRL_Berry-dos,XiaoNiu10RMP_Berry} as
\begin{equation}
\bar{D}\left( \mathbf{k}\right) \equiv 1 +\frac{e}{\hbar }%
\mathbf{\boldsymbol{\Omega }}\mathbf{\left( \mathbf{k}\right) \mathbf{%
\cdot B}}  \label{Eq:Dk}
\end{equation}%
and 2) a correction to the total band energy due to the orbital magnetic
moment~\cite{XiaoNiu10RMP_Berry}, i.e.,
\begin{equation}
\varepsilon _{M}\left( \mathbf{k}\right) =\varepsilon\left( \mathbf{k}\right) -%
\boldsymbol{m}\left( \mathbf{k}\right) \cdot \mathbf{B}\,.  \label{Eq:Em}
\end{equation}%
From the full Hamiltonian of the surface states~(\ref{Eq:H_TI}), we can
derive the following general expression for the Berry curvature
\begin{equation}
\boldsymbol{\Omega }\left( \mathbf{k}\right) =-\frac{\alpha \hbar ^{2}}{%
2\left\vert \varepsilon\left( \mathbf{k}\right) \right\vert ^{3}}\left[
\alpha \left( g\mu _{B}B_{z}-2\lambda k^{3}\cos 3\phi _{k}\right)+3\lambda
k^{2}g\mu _{B}\left( B_{x}\sin 2\phi _{k}+B_{y}\cos 2\phi _{k}\right) \right] \,\mathbf{\hat{z}} \,. \label{Eq:Omegak}
\end{equation}%
And for the two-band model, one can easily show that the orbital magnetic
moment is proportional to the product of the energy dispersion and the Berry
curvature, i.e.,
\begin{equation}
\boldsymbol{m}\left( \mathbf{k}\right) =\frac{e}{\hbar }\varepsilon(%
\mathbf{k})\boldsymbol{\Omega }\left( \mathbf{k}\right)  \label{Eq:mk}  \,,
\end{equation}%
where the energy dispersion of the upper band, $\varepsilon(\mathbf{k})$,  is given by Eq.~(%
\ref{Eq:E^s}) with $s=1$.

Equipped with Eqs.~(\ref{Eq:Dk}) - (\ref{Eq:mk}), we are ready to estimate the sizes of Berry curvature effects on the BMER. Using
the following material parameters relevant to the experiments~\cite%
{HeZhang18NP_BMER}: $\alpha =5\times 10^{5}$ m/s, $\lambda =165$ eV$\cdot $%
\AA $^{3}$, $\varepsilon _{F}=0.256$ eV, $g=2$, $B_{x}=B_{y}=B_{z}=5$ T, and
$k_{F}\sim \frac{\varepsilon _{F}}{\alpha \hbar }=7.8\times 10^{8}$ m$^{-1}\,
$, we obtain
\begin{equation*}
\max \left( \left\vert \frac{e}{\hbar }\mathbf{\boldsymbol{\Omega }\left(
\mathbf{k}\right) \mathbf{\cdot B}}\right\vert \right) \sim 0.003\ll 1
\end{equation*}%
and
\begin{equation*}
\max \left( \left\vert \boldsymbol{m}\left( \mathbf{k}\right) \cdot \mathbf{B%
}\right\vert \right) \sim 0.001\text{ eV}\ll \lambda k_{F}^{3}<\alpha \hbar
k_{F}\simeq \varepsilon _{F}  \,,
\end{equation*}%
where $\lambda k_{F}^{3}\sim 0.08$ eV. We thus conclude that the influence
of the Berry curvature on the BMER can be neglected in our present case of
interest.

\section{Summary and Conclusion}

\label{summary}
In this paper, we have developed a transport theory for a new kind of nonlinear magnetoresistance on the surface of three dimensional TIs. At variance with the UMR effect in magnetic bilayers, the nonlinear magnetoresistance does not require the presence of a conducting ferromagnetic layer and scales linearly with both the applied electric and magnetic fields; for this reason, we name it bilinear magneto-electric resistance (BMER). We have also shown that the sign and the magnitude of the BMER depends sensitively on the orientation of the current with respect to the magnetic field as well as the crystallographic axes -- a property that can be utilized to map out the spin texture of the topological surface states via simple transport measurement.

The physical origin of the BMER is a partial conversion from the nonlinear spin current to charge current in the presence of an external magnetic field. An analytical expression of the BMER is derived based on a semiclassical Boltzmann transport theory in the relaxation time approximation, which allows us to further examine various aspects of the BMER. We find that, in addition to the spin-momentum locking of the topological surface states, the cubic hexagonal warping term also plays a crucial role in generating the BMER -- the BMER vanishes in the absence of the hexagonal warping since in this case the external magnetic field can no longer deform the Fermi contour (provided that the Fermi level is not too close to the Dirac point). Also, we have shown that, the Berry curvature effect is unimportant when the Fermi level is far away from the Dirac point in which case the hexagonal warping effect is profound.

\section*{Acknowledgments}

We are grateful to Pan He, Hyunsoo Yang, Guang Bian, Axel Hoffmann, Olle Heinonen, Shufeng Zhang and Albert Fert for helpful discussions. The theoretical framework for the BMER was developed by S. S.-L. Zhang and G. Vignale at the University of Missouri and was supported by NSF Grants DMR-1406568. Detailed analysis of various aspects of the BMER as well as the manuscript preparation was done by S. S.-L. Zhang at Argonne National Laboratory and was supported by Department of Energy, Office of Science, Materials Sciences and Engineering Division through Materials Theory Institute.

\bibliographystyle{spiebib}
\bibliography{20180816_BMER-spie}


\end{document}